\documentclass[twocolumn,trackchanges]{aastex62} 

\usepackage{color}
\usepackage{lineno}
\graphicspath{{./}{figures/}}

\received{???}
\revised{???}
\accepted{???}
\shorttitle{The ridge structure of the Galactic outer disk}
\shortauthors{Yang \& Wang et al.}

\begin{document}
\title{Kinematic-Chemical analysis and Time tagging for the Diagonal Ridge Structure of the Galactic Outer Disk with LAMOST Red Giant Branch Stars}
\author{Peng Yang}
\affil{Department of Astronomy, China West Normal University, Nanchong, 637002, P.\,R.\,China}
\author[0000-0001-8459-1036]{Hai-Feng Wang}
\affil{Department of Astronomy, China West Normal University, Nanchong, 637002, P.\,R.\,China}
\affil{CREF, Centro Ricerche Enrico Fermi, Via Panisperna 89A, I-00184, Roma, Italy}
\author{Zhi-Quan Luo}
\affil{Department of Astronomy, China West Normal University, Nanchong, 637002, P.\,R.\,China}
\author{Thor Tepper-Garc\'{i}a}
\affil{Sydney Institute for Astronomy, School of Physics, University of Sydney, NSW 2006, Australia}
\affil{Centre of Excellence for All Sky Astrophysics in Three Dimensions (ASTRO-3D), Australia}
\author{Yang-Ping Luo}
\affil{Department of Astronomy, China West Normal University, Nanchong, 637002, P.\,R.\,China}

\correspondingauthor{HFW}
\email{hfwang@bao.ac.cn};\\

\begin{abstract}

We investigate the kinematic-chemical distribution of Red Giant Branch (RGB) stars from the LAMOST survey crossed matched with Gaia DR2 proper motions, and present time tagging for the well-known ridge structures (diagonal distributions for $V_R$  in the $R$, $V_\phi$ plane) in the range of Galactocentric distance $R$ = 8 to 15 kpc. We detect six ridge structures, including five ridges apparent in the radial velocity distribution and three ridges apparent in the vertical velocity, the sensitive time of which to the perturbations are from young population (0$-$3 Gyr) to old population (9$-$14 Gyr). Based on an analysis of the evolution of angular momentum distribution, we find that four ridges are relatively stationary, while another is evolving with time, which is confirmed by the difference analysis at different populations and supporting that there might be two kinds of dynamical origins. Furthermore, ridge features are also vividly present in the chemical properties ([Fe/H], [$\alpha$/Fe]). The comparison between the north and south hemispheres of the Galaxy does show some differences and the ridge features are asymmetrical. Moreover, we find that diagonal ridge structures may affect the shape of the rotation curve, which is manifested as fluctuations and undulations on top of a smooth profile. Finally we speculate that the bar dynamics should be not enough to explain all ridge properties including the break feature in the $V_Z$-$L_Z$ plane.

\end{abstract}


\section{Introduction} 
Galactic seismology is the inference of the Galactic potential and Galactic sub-structure from the dynamical analysis of the observed perturbations in the Milky Way \citep{2017IAUS..321..108C,Bland2021}\footnote{For a brief history, see the introduction part of \citet{Bland2021}.}. It is also characterising the Milky Way that is non-equilibrium and non-stationary in the asymmetric potential, and will undoubtedly help us have a more comprehensive understanding for the dynamic origins and evolution of the Galaxy \citep{Widrow2012, Carlin2013,Widrow2014,Liu2018,2018Natur.563...85H,Belokurov2018,Wang2018a,Wang2018b,wang2019,Wang2020a,Wang2020b,Wang2020c,wang2022a,wang2022b,wang2022c,Lopez2019, Lopez2019review,Lopez2020,Trick2019,Yu2021,Tepper2021,Tepper2022,2022ApJ...936..103G}.

The cornerstone Gaia mission \citep{Gaia2016} provides the most accurate proper motions and photometry to date. The G-band stars in Gaia DR2\footnote{Note that this work was finished before Gaia DR3 and for the range 8$-$15 \,kpc we are concerned about, DR2 and DR3 will have no large difference.} is up to 1.69 billion, among them, 1.33 billion stars have 5 parameters, 360 million stars have 2 parameters, and 7.22 million stars have radial velocity information \citep{Gaia2016,Gaia2018b}. With the help of Gaia data, many exciting discoveries hidden in the Milky Way Galaxy are revealed in recent years. As described in \citet{Antoja2018}, our Galactic disc is phase mixing shown as the ``snails" and ``ridge" features, both of which are morphology distribution in the phase-space.

The ``Ridge" structure, which describes the stellar radial velocity distributions in the plane of radial distance ($R$) and azimuthal velocity ($V_\phi$) with many diagonal features. It was first revealed in the solar neighborhood by \citet{Antoja2018} using Gaia DR2 data, along with many other substructures such as arches and shells. Meanwhile, \citet{Ramos2018} suggested that the moving groups and arches in the U$-$V velocity space of the solar neighborhood are projections of the diagonal ridges.

In addition, \citet{Khanna2019} has revealed the ridge pattern for the vertical distance ($Z$), vertical velocity ($V_Z$), radial velocity ($V_R$), metallicity ([Fe/H]), and abundance ([$\alpha$/Fe]) with GALAH southern sky survey \citep{2015MNRAS.449.2604D}. The ridge features in the rotation velocity distribution were also unravelled in \citet{Kawata2018} and they suggested that the ridge are linked to the Galactic bar and spiral arm in the range of 5 to 12\,kpc. Recently, the ridge in the outer disc of the Milky Way is unveiled in \citet{Antoja2021} using the data of Gaia EDR3 without radial velocity and age information. \citet{Bernet2022} also find some moving groups are existing in the diagonal lines of the $R$$-$$V_{\phi}$ plane, which are matched with some known diagonal ridges. And the locations of the diagonal ridges can be considered as continuous manifolds in the 6D phase-space, thus then the diagonal ridges are projections of these manifolds.

Possible links between the Milky Way spiral arms, ridges, and moving groups in local space are unravelled in \citet{Khoperskov2022}, they found that stars in angular momentum over-densities can track density ridges over many \,kpc with the help of high-resolution spiral galaxy simulation as well as Gaia DR2 and EDR3 data. \citet{Recio2022} have also found that most diagonal ridges are metal-rich using metallicity in Gaia DR3 RGB stars without age information. And they also identified seven ridge structures in the chemistry. 

Using a toy model, \citet{Antoja2018} found that the diagonal ridge structure could be generated by phase mixing from an out of equilibrium state corresponding to the perturbation from the pericentric passage of the Sagittarius dwarf galaxy (see also \citet{2022arXiv220603495A}), but which could also be strongly affected by the Galactic bar and/or spiral structure. \citet{Hunt2018} found that a test particle model with winding spiral structure could reproduce the observed diagonal ridges well. \citet{Fragkoudi2019} used collision free N-body simulations and also combined with orbital integration to find that the ridge structure is caused by the outer lindblad resonance (OLR) of the Galactic bar. \citet{Barros2020} found that the stellar orbit captured by spiral resonance can also produce ridge characteristics and meanwhile, \citet{Laporte2020} found that Galactic bar plays important roles both in the ridge evolution and shaping the Galactic disc. Moreover, \citet{Monari2019a,Monari2019b} suggested that no less than six ridges in the local action space are related to the resonance of the bar using a slow rotating Galactic bar model with $\Omega$ = 39 km s$^{-1}$ kpc$^{-1}$. 

In general, for the mechanisms of the snails and ridges, the external perturbation such like the Sagittarius dwarf galaxy interaction with the Milky Way \citep{Antoja2018,Binney2018,2019MNRAS.485.3134L,Bland2021,Craig2021}, the internal dynamics without external disturbance such like spiral arms, outer lindblad resonance of the bar \citep{Kawata2018,Monari2019a,Khoperskov2019,Barros2020}, the coupling mechanisms of spiral arms and bars or Sagittarius perturbation simulations for ridges are shown in \citet{Khanna2019,Khoperskov2022}. To date, whether or not these asymmetric patterns are from internal or external or both mechanisms is still unclear.

Using Main-Sequence-Turn-Off and OB type stars selected from the LAMOST Galactic spectroscopic surveys  \citep{Deng2012}, \citet{Wang2020c} have reconstructed the ridge pattern in the chemo-kinematical space. They revealed three ridges which are showing some of chemistry patterns, young populations feature and north-south comparisons, two are relatively stable but one is changeable with age, implying there might have two kinds of ridge patterns with different dynamical origins. However, in \citet{Wang2020c}, they only focus on the range within 12 \,kpc and found only three ridges, and they are lack of detailed quantitative north-south comparisons with temporal evolution. Motivated by this, during this paper we will make full use of RGB stars to explore further about the ridge topic mainly from the observational point of view.

The paper is structured as follows: in Section 2, we describe the dataset we adopt, sample selection and coordinate transformation in this work; in Section 3 we present our main ridge patterns for kinematic, chemical, dynamical distributions, and discussion for some possible scenarios qualitatively, in particular for the bar; in Section 4, we summarise the results in general and make a prospect.

\begin{figure}
  \centering
  \includegraphics[width=0.45\textwidth]{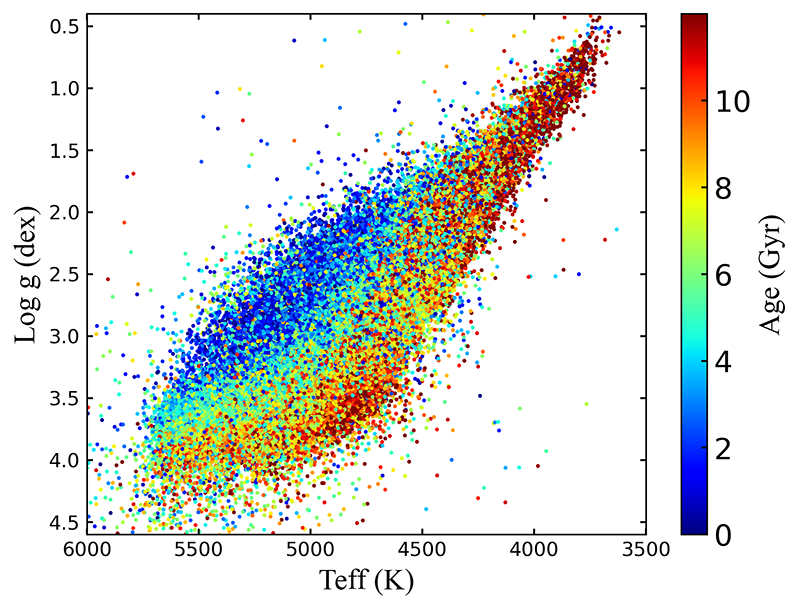}
   \centering
  \includegraphics[width=0.475\textwidth]{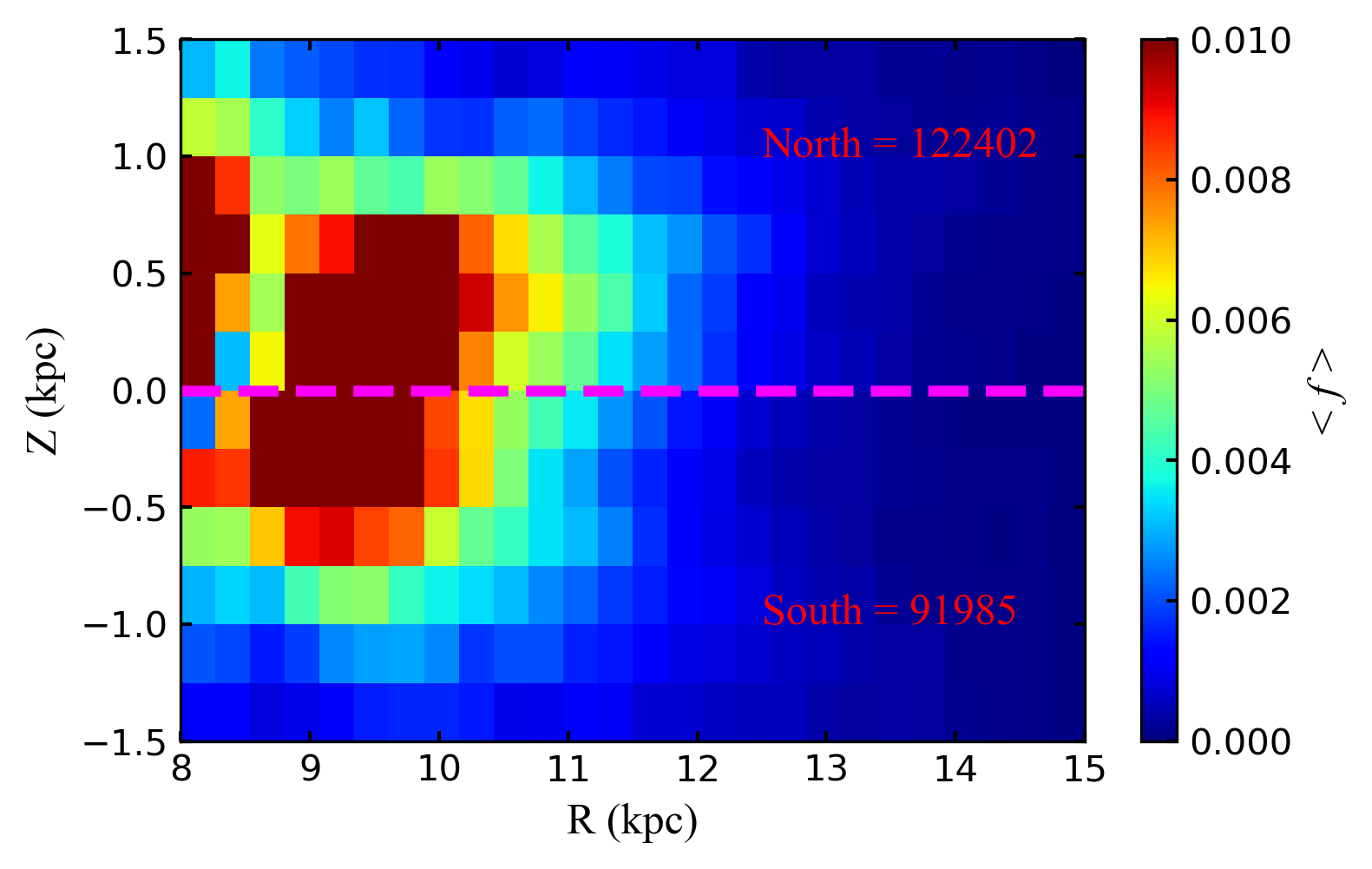}
    \centering
  \includegraphics[width=0.45\textwidth]{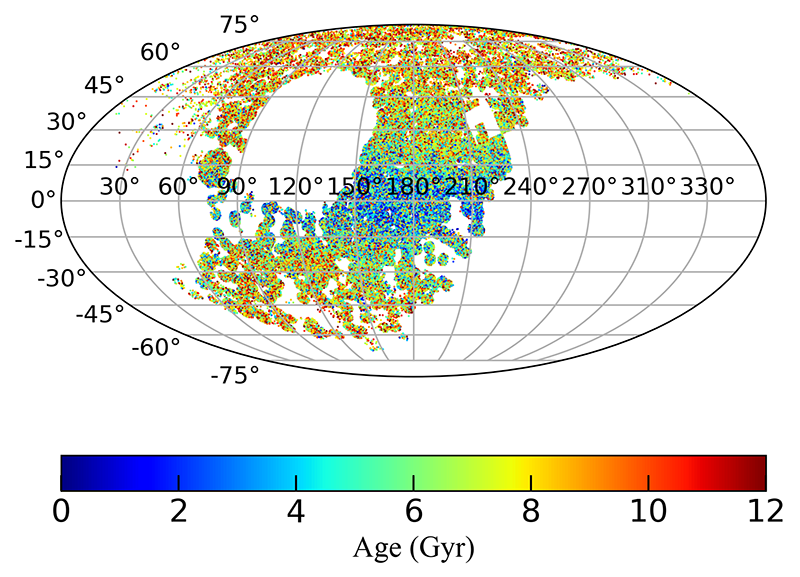}
  \caption{Top panel: distribution of RGB stars used in this work on the Teff$-$log g plane coloured by age; Middle one: the density distribution of RGB stars in the $R, Z$ plane and the number of stars in the north of the Galactic disc (122402) is more than that in the South (91985), which is normalised by the total number; Bottom panel: distribution of RGB stars in the Galactic longitude and latitude coordinate system.Younger stars are nearer the low galactic latitude naturally, in the meanwhile older stars are mainly distributed in the high galactic latitude.}
  \label{sample1}
\end{figure}

\section{Data} 

Age and mass of the sample we use are from the red giant branch stars of LAMOST DR4 estimated by \citet{Wu2019} with kernel principal component analysis (KPCA). Then we cross match the catalog with \citet{xiang2019} for metallicity, abundance with Data–Driven $Payne$ method \citep{Ting2019}, and the distance in this work is based on the bayesian estimation method from \citet{Carlin2015} for LAMOST DR5, which is also calibrated by \citet{xuyan2020}. In general, the catalog contains 640986 RGB stars, which has the radial velocity, age, mass, metallicity, chemical abundance and distance we need. The radial velocity uncertainty is 5 km s$^{-1}$ estimated by using the LAMOST stellar parameter pipeline of Peking University (LSP3) \citep{xiang2017a}, the error of mass and age determined by the KPCA  is 10\% and 30\% respectively.  Metallicity error is 0.1 \,dex,  [$\alpha$/Fe] error is about 0.05 dex (which are also wildly used in our previous work (\citet{2022arXiv220506144L}) and the distance uncertainty is about 15\%. The proper motions information of the sample comes from the Gaia DR2 catalog, and the uncertainties of the proper motion is 0.06 mas yr$^{-1}$ (for G \textless 15 mag), 0.2 mas yr$^{-1}$ (for G = 17 mag) and 1.2 mas yr$^{-1}$ (for G = 20 mag) \citep{Gaia2018b}.

\begin{figure*}
  \centering
  \includegraphics[width=0.9\textwidth,height=0.3\textwidth]{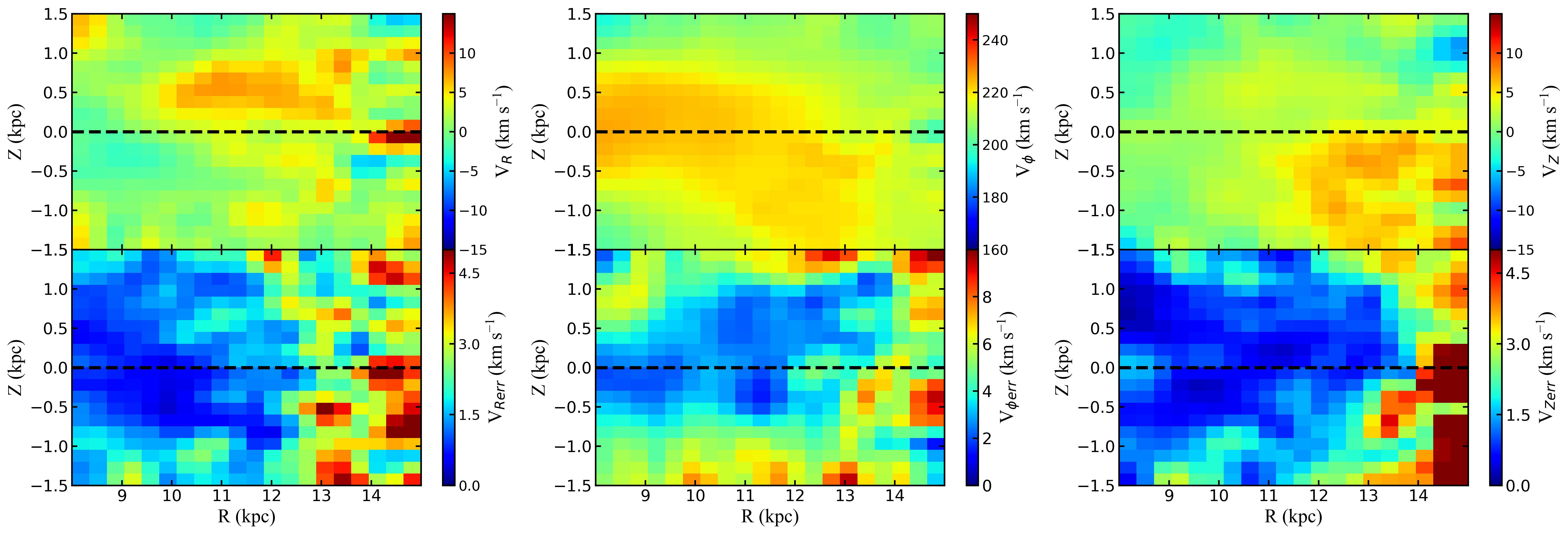}
  \caption{Three-dimensional velocity distributions of the final sample in the $R-Z$ plane, the top three are $V_{R}$, $V_{\phi}$ and $V_{Z}$ velocity and the error is determined by bootstrap method in the bottom three. $R$ and $Z$ are limited to [8,15] \,kpc and [$-$1.5,1.5] \,kpc respectively.}
  \label{R-Z-V}
\end{figure*}

\begin{figure*}
  \centering
  \includegraphics[width=0.9\textwidth,height=0.7\textwidth]{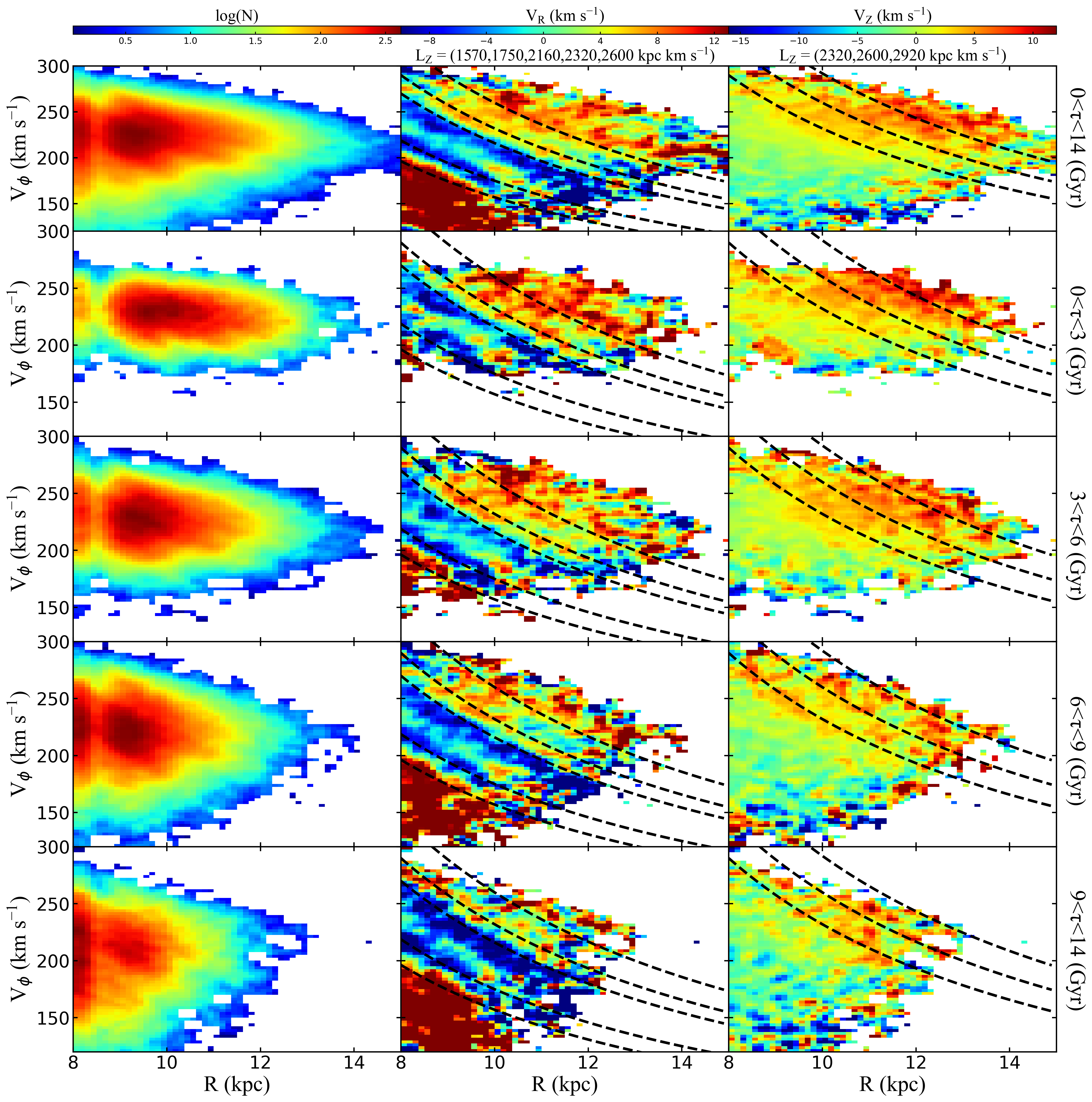}
  \caption{The left column shows the star counts distribution in log scale, the middle column shows the radial velocity $V_{R}$ distribution, and the right column shows the vertical velocity $V_Z$ distribution in different age populations. The black dashed lines in the radial velocity diagram represent the constant angular momentum $L_Z$ = (1570, 1750, 2160, 2320, 2600) kpc km s$^{-1}$.The black dashed lines in the right vertical velocity diagram represent the constant angular momentum $L_Z$ = (2320, 2600, 2920) kpc km s$^{-1}$.}
  \label{f-Vr-Vz}
\end{figure*}

After the cross match for LAMOST and Gaia dataset, we use the following criteria to select the final sample to present our results:

(1) 8 \textless $R$ \textless 15 \,kpc and $-$1.5 \textless $Z$ \textless 1.5 \,kpc;

(2) SNR \textgreater 10;

(3) 0 \textless Age \textless 14\,Gyr;

(4) parallax \textgreater  0;

(5) 50 \textless $V_\phi$ (km s$^{-1}$) \textless 300 ;

\begin{figure*}
  \centering
  \includegraphics[width=0.8\textwidth,height=0.8\textwidth]{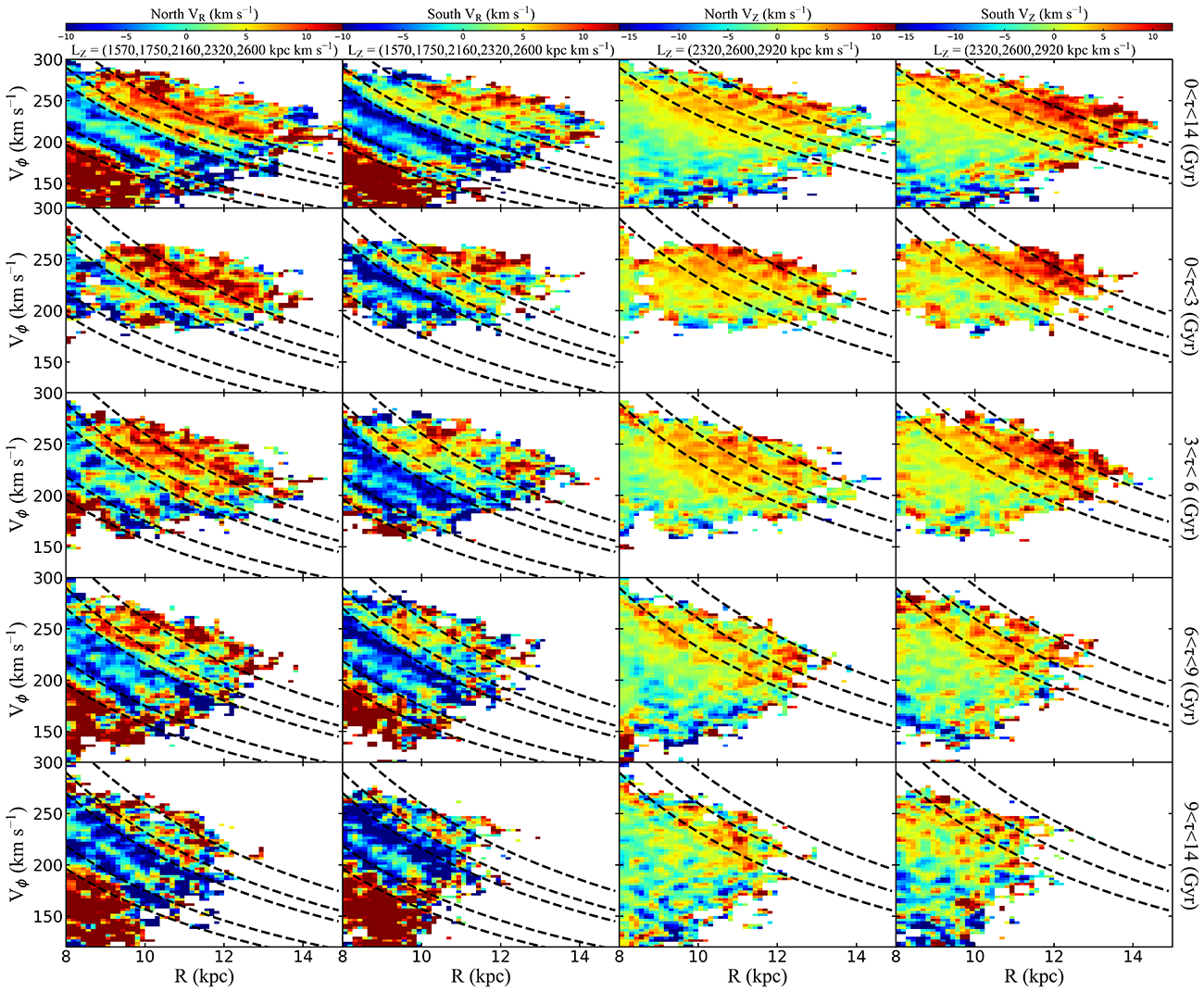}
  \caption{The left two columns show the radial velocity distribution on the north and south sides of the Galactic disk, and the right two columns show the vertical velocity features. The north and south sides of the Galactic disk have obvious ridge pattern and there are clear asymmetries indicated by the different colour distributions.}
  \label{R-vphi-NSvR}
\end{figure*}

\begin{figure*}
  \centering
  \includegraphics[width=0.8\textwidth,height=0.8\textwidth]{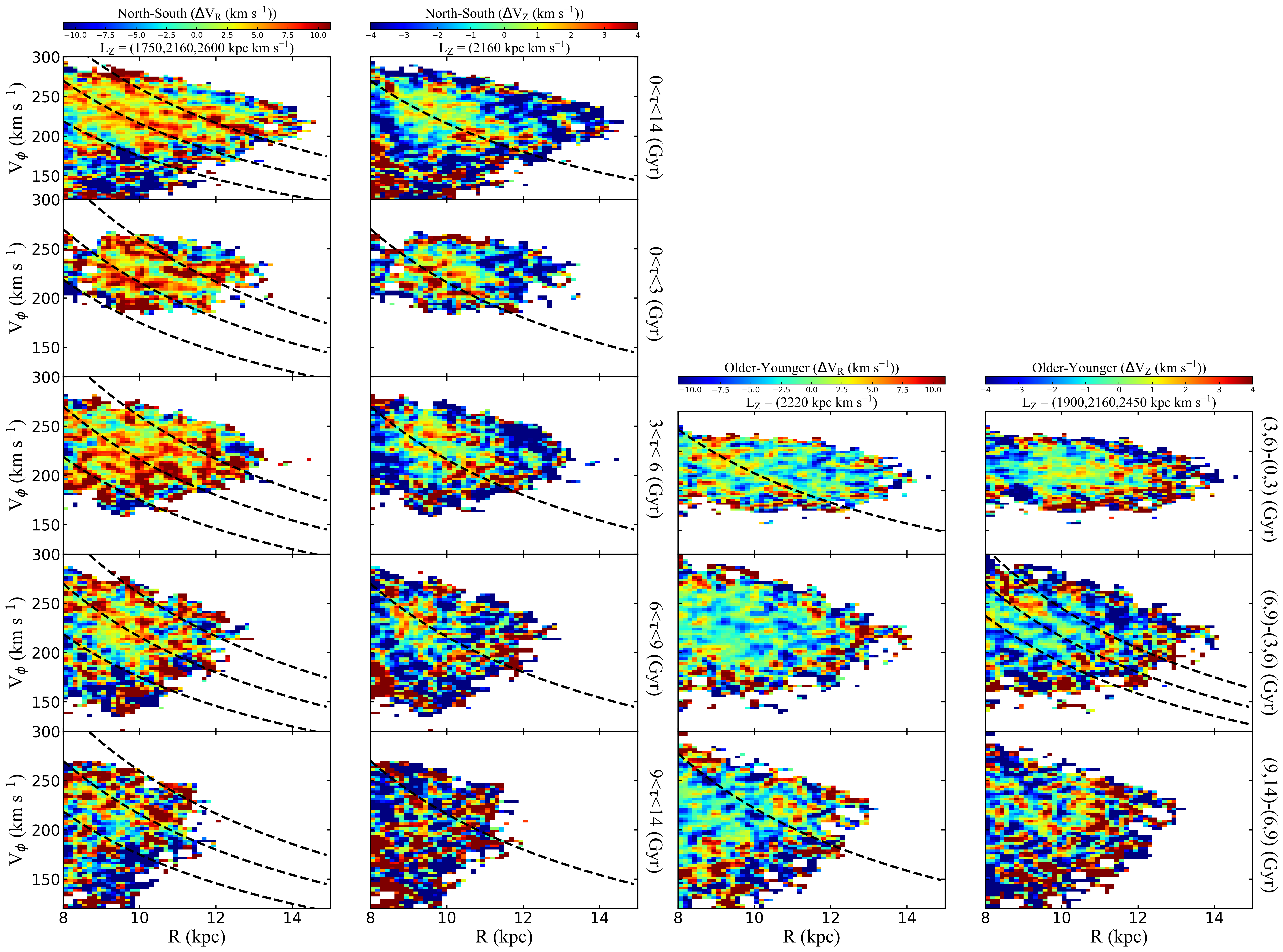}
  \caption{The left two columns show the radial velocity (black dashed lines $L_Z$ =1750, 2160, 2600 kpc km s$^{-1}$) and vertical velocity ($L_Z$ is 2160 kpc km s$^{-1}$) difference distributions of the north minus south sides. There are still three ridges in the same location as Fig. \ref{f-Vr-Vz}. The amplitude of the south and north ridges is different, which is showing the clear asymmetries, especially for the vertical non-null values. The right two columns show the velocity differences at different age populations for velocity ($L_Z$ is 2220 kpc km s$^{-1}$ for $V_R$ , 1900, 2160, 2450 kpc km s$^{-1}$ for $V_Z$), which are showing there are remaining one and three time-evolving ridges respectively.} 
  \label{NSvdifference}
\end{figure*}

Then we have 214387 RGB stars, the age distribution on the Teff$-$log g plane, star counts distribution on the $R$$-$$Z$ plane, age distribution on the celestial coordinate are shown in Fig. \ref{sample1}, all of which are showing the sample properties and parameter distributions. Fig. \ref{R-Z-V} shows the 3 dimensional velocity distribution (top three) with bootstrap error analysis (bottom three), the velocity substructure around 10$-$11 \,kpc in the left one,  the asymmetric drift in the middle one and the vertical bulk motions in the right panel are clearly shown here, which are consistent with the results in \citet{Wang2018a,Wang2020a,Gaia2018b}.

\begin{figure*}
  \centering
  \includegraphics[width=0.9\textwidth,height=0.6\textwidth]{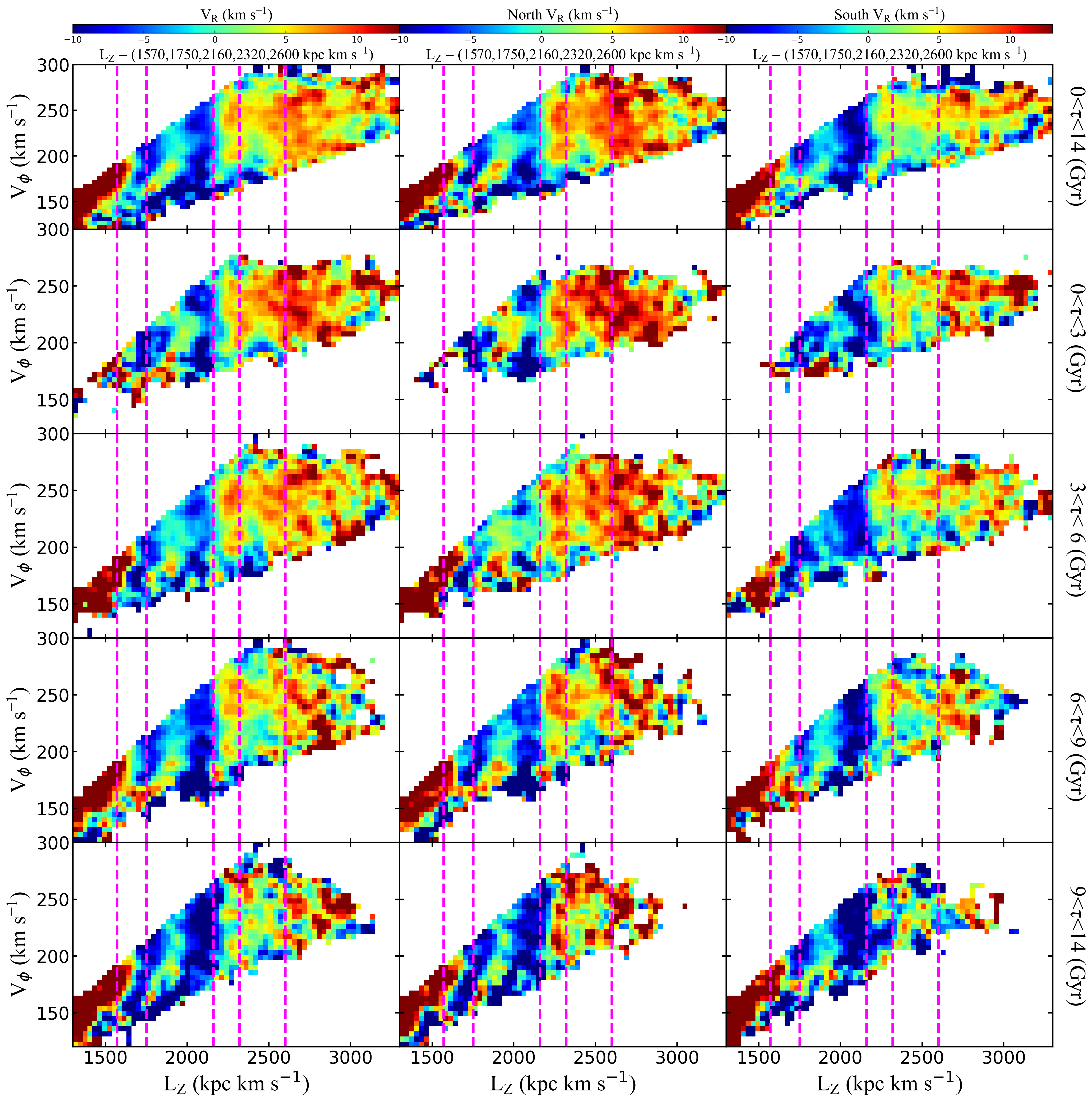}
  \caption{The radial velocity distribution of RGB stars used in this paper are shown in the $L_Z$ and $V_\phi$ plane. The five vertical fuchsia lines in the figure represent constant angular momentum $L_Z$ = (1570, 1750, 2160, 2320, 2600) kpc km s$^{-1}$. We can see that third ridge (R3 in the context) changes with age or has an offset, others are relatively stable in the left column. The middle and right columns show the distribution of sample on the north and south sides, which are similar to the left one but the north-south differences are detected.}
  \label{Lz-vphi-Vr}
\end{figure*}

Notice that the position of the Sun $R_{\odot}$  = 8.34 kpc \citep{Reid2014}, and its vertical distance to the disk is $Z_{\odot}$ = 0.027 \,kpc \citep{Chen01}. The local standard of rest velocity in the solar neighborhood is 238 km s$^{-1}$\citep{Schonrich12}. The solar motions are [$U_{\odot}$, $V_{\odot}$, $W_{\odot}$]= [9.58, 10.5, 7.01] km s$^{-1}$ \citep{Tian2015}, different solar motions will not change our final conclusions in this work. With the help of $Galpy$ \citep{Bovy2015}, we present the kinematics results as follows. We have adopted a right-handed Galactocentric cartesian coordinates with $X$ increasing outward from the Galactic centre, $Y$ in the direction of rotation, and $Z$ positive towards the North Galactic Pole (NGP). Cylindrical velocities $V_R$, $V_\phi$, and $V_Z$ are defined as positive with increasing $R$, $\phi$, and $Z$.

\begin{figure*}
  \centering
  \includegraphics[width=0.8\textwidth,height=0.4\textwidth]{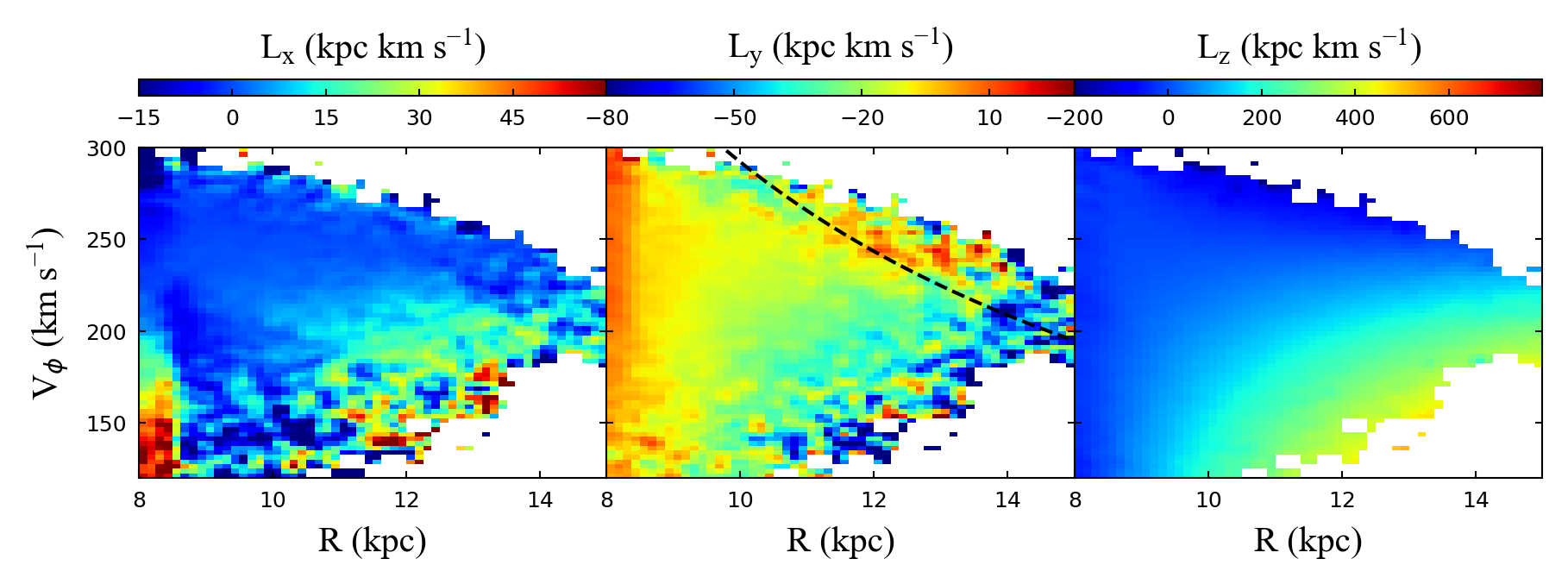}
  \caption{Distribution of 3D angular momentum component in the $R$-$V_\phi$ plane: angular momentum in the x direction (left), y direction (middle) and z direction (right). The only one dashed line represents the constant angular momentum curve implying there might be a ridge corresponding to the previous R6 = 2920 kpc km s$^{-1}$.}
  \label{Lxyz}
\end{figure*}

\section{Results and Discussion} 

\subsection{Ridge distribution in kinematical and dynamical space}

Fig. \ref{f-Vr-Vz} presents the star counts ($log(N)$), radial velocity ($V_{R}$), and vertical velocity ($V_{Z}$) on the $R$ and $V_\phi$ plane in different age populations. The top panel shows the entire sample (0$-$14\,Gyr), and the bottom four panels show ridge structures for populations of different ages which are marked in the extreme right. The constant angular momentum curve values are denoted on the top of the columns. The values of these curves in the $V_R$ diagrams are 1570, 1750, 2160, 2320, 2600 kpc km s$^{-1}$ respectively, and the values in the $V_Z$ diagrams are 2320, 2600 and 2920 \,kpc km s$^{-1}$ respectively. It can be seen clearly there are also some ridge patterns in the density distribution similar to \citet{Antoja2018}. As shown in the middle column for $V_R$, it is clearly presented that there are five ridges from top one to bottom one, which are matched well with the constant angular momentum dashed curved lines except the third one, it appears that there is an offset with temporal evolution. The angular momentum shift in the ridge were also mentioned in the modelling work of \citet{2019MNRAS.490.5414F}

In addition, we find three diagonal ridge structures in the $V_Z$ distributions in the right column, one of which ($L_Z$=2920) is corresponding to the new ridge discovered in the \citet{Antoja2021} ($L_Z$=2750) in consideration of the solar values, the relative difference of the $L_Z$/$L_\sun$ is only 4\%. So we reveal six ridge structures in the radial velocity diagram ($V_{R}$) or vertical velocity diagram ($V_{Z}$), and present the time stamps on it. These ridges are showing from young population 0$-$3 Gyr to old population 9$-$14 Gyr, so the sensitive time of the ridge to the possible dynamical perturbations might be from 0$-$14 Gyr. Since younger populations could be tracer of the early Milky Way and older populations could be tracer of the late Milky Way,  perhaps we could conjecture that the ridge is a long-lived structure.

\begin{figure*}
  \centering
  \includegraphics[width=0.8\textwidth,height=0.8\textwidth]{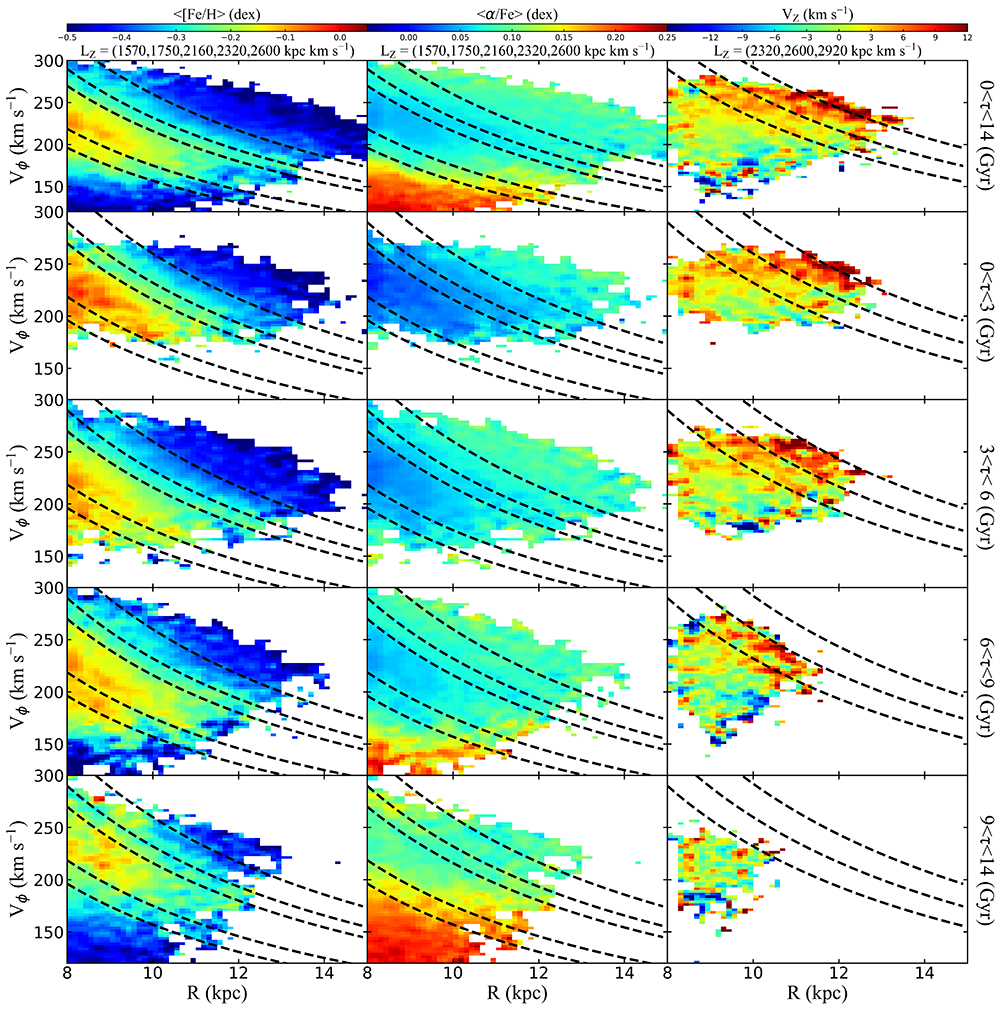}
  \caption{The left and middle columns show the metallicity and chemical abundance distribution corresponding to the vertical height of [$-$1.5, 1.5] \,kpc respectively. The right one shows the vertical velocity distribution but with vertical height of [$-$0.2, 0.2] \,kpc. Obviously the ridge signals are detected in all three panels, although it is not as clear as the kinematic space, the number of ridge here is also less than that in the kinematic space.} 
  \label{R-vphi-element}
\end{figure*}

To investigate more about the ridge asymmetrical features, we divide our sample into the north and south sides as shown in Fig. \ref{R-vphi-NSvR}. As seen clearly, we can still detect clear ridge signals, and there is no very strong asymmetry for constant angular momentum curves, but other differences are discovered. It does not come at surprise for some differences, since we know the azimuthal velocity asymmetries or asymmetric drift is existed. For example, we can see the first red ridge (2600) in the north are clearer and stronger than that in the south for the radial velocity, but for the vertical velocity, the south distribution is clearer and wilder than the north. The third blue ridge (2160) clearly presents in the south but not so strong in the north. In short, although the location has no large difference but it doesn't mean the ridge is symmetrical, more quantitative comparisons using north minus south are shown in Fig. \ref{NSvdifference}.

Interestingly, as shown in Fig. \ref{NSvdifference}, the left two panels show the radial velocity and vertical velocity difference distributions of the north minus south sides. In the $V_R$ diagram of Fig. \ref{NSvdifference}, three diagonal ridges have the same position as those in Fig. \ref{f-Vr-Vz}, and one diagonal ridge has the same position in the $V_Z$ diagram. Notice the amplitudes of the northern and southern diagonal ridges, expressed as colour velocity differences, should differ. The right two columns show the velocity differences at different age populations for velocity (e.g., population in [3$-$6] \,Gyr minus population in [0$-$3] Gyr), it shows that there are one and three unstable time-evolving diagonal ridges in $V_R$ and $V_Z$ respectively. In consideration of the radial motions and moving groups in the north side as mentioned in \citet{Wang2020a}, the radial velocity difference in the left column might imply more coupling kinematic properties. Anyway, we have detected the ridge asymmetries in particular the vertical motions.

All these are consistent with the implications that the disk kinematics is asymmetrical, ridge is also asymmetrical in the phase space location especially the amplitude. Moreover, there are two kinds of ridges which have stable or nonstable evolution with time. Notice that some dashed lines have slight difference referenced to Fig. \ref{f-Vr-Vz} ($L_Z$ = (1570, 1750, 2160, 2320, 2600) kpc km s$^{-1}$) during this paper, all the dashed lines value (constant angular momentum) difference is within 9\% so it will not change our any conclusions shown in the later parts. 

\begin{figure*}
  \centering
  \includegraphics[width=0.8\textwidth,height=0.8\textwidth]{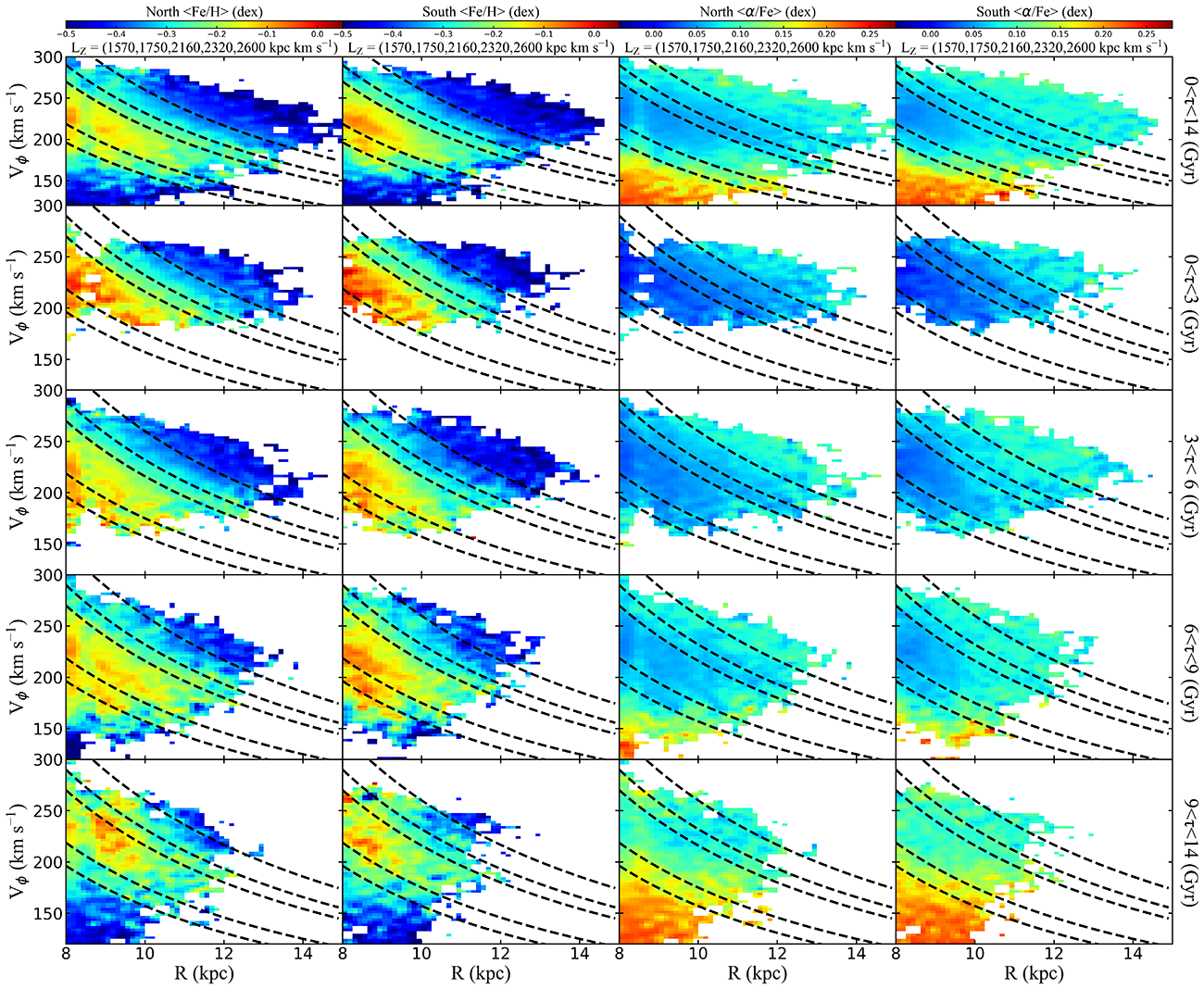}
  \caption{Similar to Fig. \ref{R-vphi-element} for the chemical analysis but the sample is divided into north and south sides.}
  \label{R-vphi-NSelement}
\end{figure*}

In order to better present the six diagonal ridges in Fig. \ref{f-Vr-Vz} and subsequent figures, the diagonal ridges for $L_Z$=1570, 1750, 2160, 2320 and 2600 kpc km s$^{-1}$ in $V_R$ diagram are denoted as R1, R2, R3, R4 and R5 respectively, and the diagonal ridges with $L_Z$=2920 kpc km s$^{-1}$ in $V_Z$ diagram are denoted as R6. The R1, R2, R4, R5 and R6 ridges are relatively stable and R3 is evolving with time, which is manifesting there are two kinds of ridges with different kinds of origins as mentioned. 

Furthermore, we then make more analysis using angular momentum, Fig. \ref{Lz-vphi-Vr} shows the ridge features in the $L_Z$ and $V_\phi$ plane coloured by the radial velocity. By focusing on the five ridges on the left column we could clearly see the third ridge has fluctuations and the others keep relatively stable, this  also confirms two kinds of dynamical mechanisms again. What's more, the north and south distribution is also displayed in the middle and right columns, the ridge features marked by the fuchsia dashed lines. It is displaying the north-south difference especially for the third and fifth one, by focusing on the third row panels we can see the north third ridge is weaker than that in the north and the fifth ridge is reversed. Notice that the velocity error of  2$-$4 km s$^{-1}$ and distance error of 10$-$15$\%$  will not change the physical conclusions about ridge in this work.

We also have also analyzed the distribution of 3D angular momentum of RGB stars in the $R$-$V_\phi$ plane. It is shown that the $L_x$ and $L_z$ has increasing trend along with $R$, in contrast to the $L_y$, which has decreasing trend in Fig. \ref{Lxyz}. There are almost no ridge features except the middle panel, which has ridge signal shown as the black dashed lines corresponding to the R6 ridge. 

\subsection{Ridge distribution with chemistry}

In this part we mainly present the analysis about chemistry for this structure. As shown in Fig. \ref{R-vphi-element}, the left two panels are the metallicity ([Fe/H]) and element abundance ([$\alpha$/Fe]) pattern in the range of vertical height of $-$1.5 to 1.5 \,kpc. We could see two ridges indicated by the red in the fourth subfigure of the left diagram and one ridge indicated by the blue in the top subfigure of the middle column respectively. As seen the ridge is mainly consisted of metal rich and $\alpha$ poor stellar population, the metallicity is about $-$0.05 dex and [$\alpha$/Fe] is about 0.03 dex, which might imply the ridges are mainly existing in the thin disk. The disk chemical gradient patterns along with the azimuthal velocity and distance are also showing in the range $V_\phi$ larger than 200 km s$^{-1}$, which is corresponding to the negative ([Fe/H]) and positive gradient ([$\alpha$/Fe]) for the overall trend.

Here we also show the vertical velocity distribution in the narrower vertical range ([$-$0.2, 0.2] kpc) as \citet{Khanna2019}, three ridges shown in the top are found during the analysis, which are a complement for the Sec. 3.1. Moreover, the south and north comparison for ridge traced by chemistry is shown in Fig. \ref{R-vphi-NSelement}, it seems that the metallicity distribution in the south is clearer than that in the north but the $\alpha$ pattern is clearer in the north, so it might imply that the ridge is not one-to-one correlation.

\begin{figure*}
  \centering
  \includegraphics[width=0.8\textwidth,height=0.8\textwidth]{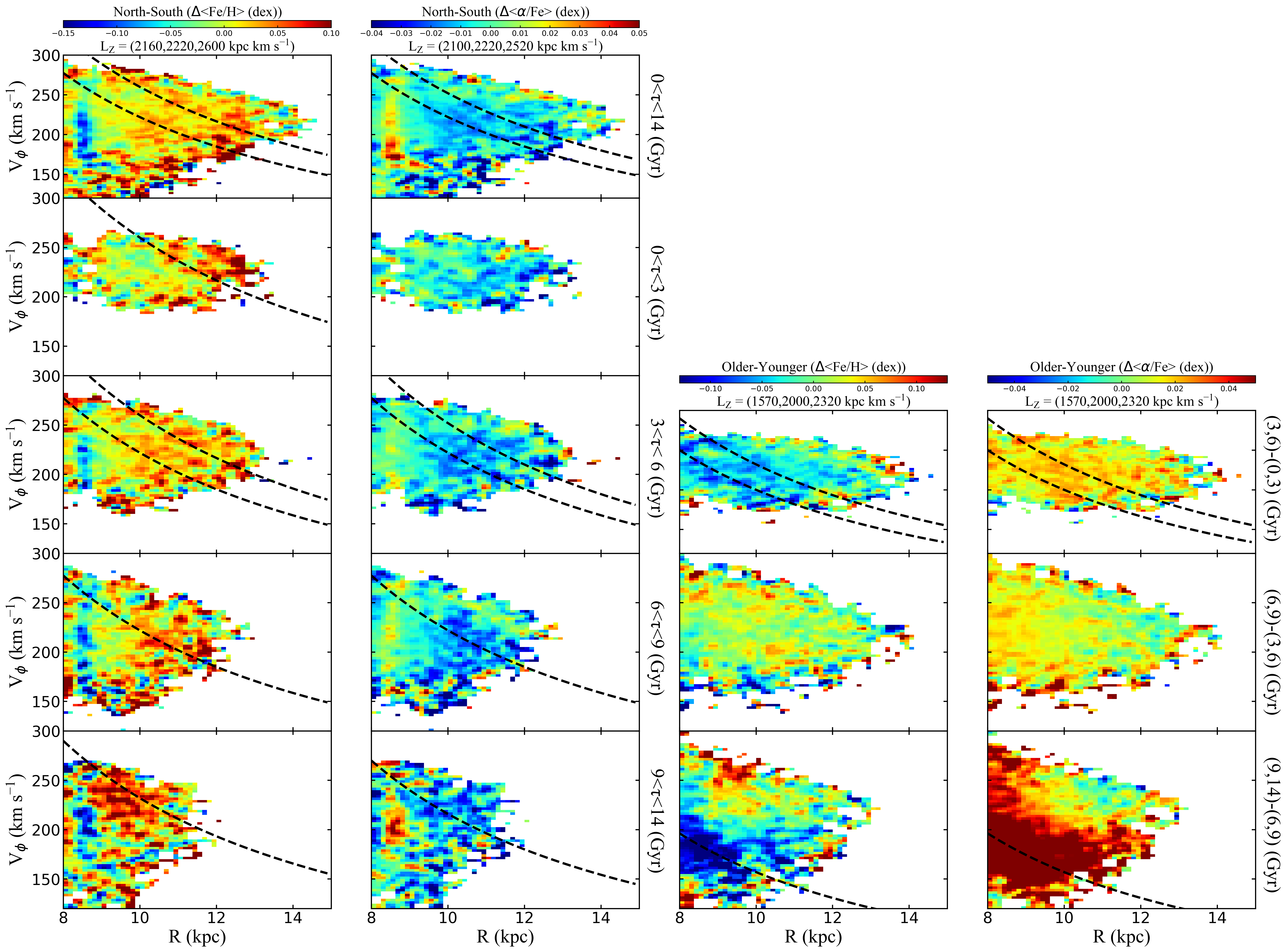}
  \caption{Similar to Fig. \ref{NSvdifference} it shows the chemical difference in the north and south sides (left two columns), and meanwhile, the difference at different age populations (right two columns). Three ridges signals are detected in the [Fe/H] and [$\alpha$/Fe] in the left two columns. Three ridges signals are detected in the right two columns, as shown the dashed lines, which are slightly different from the velocity analysis. It is also clearly shown that north-south has chemical difference for ridges.}
  \label{NSelementdiiference}
\end{figure*}

Similar to Fig. \ref{NSvdifference}, in Fig. \ref{NSelementdiiference} we analyze the chemical difference in the north and south sides (left two columns), and meanwhile, the difference at different age populations (right two columns). Three ridges signal are detected in the [Fe/H] and [$\alpha$/Fe] in the left two columns, and three time-evolving ridges are found in the right two columns, as shown the dashed lines, again, small differences for the angular momentum curves do not mean different ridges. Here we try not to over-inteprate the chemical differences, but the non-zero values in the chemistry clearly shows that the ridge has asymmetries for chemistries in the north and south sides. To our knowledge, this is the first time to uncover this intriguing phenomenon.

\begin{figure*}
  \centering
  \includegraphics[width=0.8\textwidth,height=0.8\textwidth]{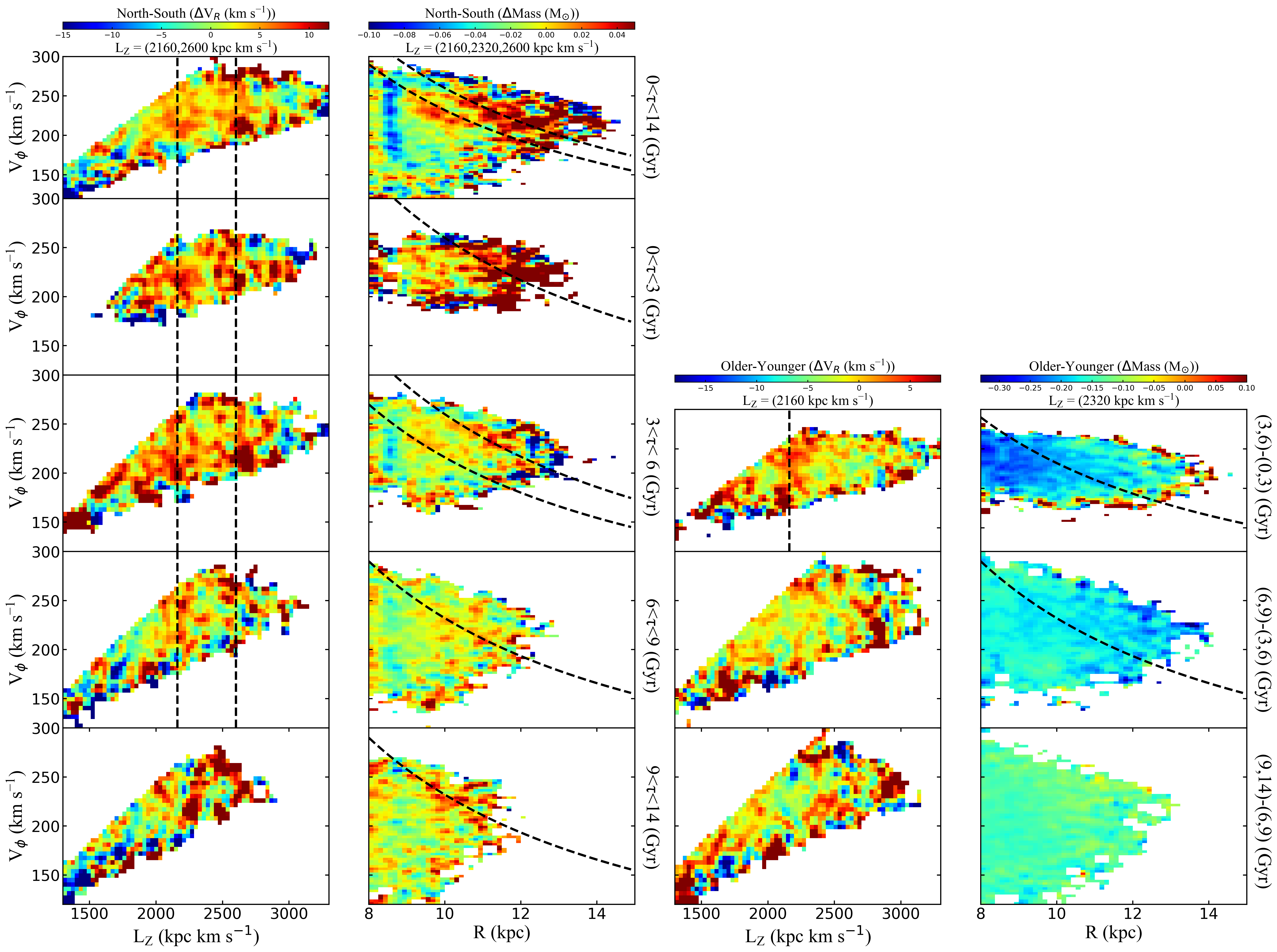}
  \caption{The left two columns are radial velocity and mass difference distribution of north minus south sides in the $R$ (or $L_Z$) and $V_\phi$ plane, right two columns are the ridge difference at different age populations which are helpful for one to find time-evolving ridge.} 
  \label{NS(MassLz)difference}
\end{figure*}

Finally and similarly, Fig. \ref{NS(MassLz)difference} displays mass difference of the north and south sides in the $R$ and $V_\phi$ plane, and also radial velocity which is as complement of Fig. \ref{Lz-vphi-Vr} in azimuthal velocity and angular momentum plane (left two panels). Meanwhile, the difference at different age populations are also explored here (right two columns) in order to investigate more asymmetrical properties. It is showing that at least two ridges are found in the first column. In the meantime, there are also asymmetries for ridge indicated by the stellar mass distribution of north minus south in the $R$ and $V_\phi$ plane. The ridge signal here is weaker than that in the kinematics, implying the sampling and mass precision need to be improved in the future work. Moreover, we also note that there is one time-evolving ridge left in the third and the final column respectively.

\subsection{Discussion}

\subsubsection{More comparisons and the bar contribution}

\citet{Antoja2021} discovered abundant ridge structures in the outer disc of the Milky Way using the data of Gaia EDR3. However, they did not investigate more details using radial velocity, chemistry, mass and age populations for the ridge, which are explored in this work. \citet{Recio2022} have also adopted RGB stars from Gaia DR3 to investigate the distribution of diagonal ridges in chemical space and found seven diagonal ridges using general sample in the actions space, it seems that four of which are significant in their Figure 31. But the volume of the disk probed here is a bit farther for chemistry compared to the recent DR3 work and we are using the new chemistry information like [$\alpha$/Fe] and ages in this work, then present the north-south asymmetries analysis for the ridge.

By focusing on the ratio of $L_Z$/$L_\sun$, when comparing with more recent works in consideration of solar radius and values. We find some ridges in \citet{Ramos2018} can be expressed as: L2 (1.148), Sirius (1.019), Hyade (0.935), Hercules (0.808), L11 (0.740), L12 (0.668), L16 (1.185). More others like \citet{Bernet2022}, the ratio of the Arch/Hat should be 1.18, and the ratio of the new ridge in \citet{Antoja2021} is 1.353. In our work, six ridges ratios are: 0.758, 0.844, 1.042, 1.119, 1.255, and 1.409, respectively. If we compare our six ridges carefully, it can be seen they might be corresponding to the L11, Herclues, Sirius, Arch/Hat, L16, new ridge of \citet{Antoja2021}.

\citet{Wang2020c} investigated the chemo-dynamical evolution of the ridge structure of the Galactic disc. They found three ridge structures from young population to old population and discovered one type is relatively stable with age, the other type changes with time. Then, they attempted to relate the observational results to the phase mixing of the spiral arm of the Milky Way and the disturbance of the Sagittarius dwarf spheroidal galaxy (Sgr), this coupling mechanisms could also be found more details in \citet{Khanna2019}. Using Gaia DR2 and GALAH DR2 Southern Sky Survey, \citet{Khanna2019} have found clear ridge features ($L_Z$ = (1350, 1600, 1800, 2080)). More importantly, they also used test particle simulations and N-body simulations to explore the origins of the ridge features, their results show that in order to unify the ridge features with different kinds of ridges and vertical motions, both the spiral arms phase mixing and the interaction between the Sagittarius dwarf galaxy with Milky Way might be better to advance our understanding for the ridges and snails. To be more specific, the spiral arm mechanisms in their model could reproduce different kinds of ridges with different energy, when considering the Sgr perturbation the in-plane and vertical motions could be reproduced well simultaneously. All these show the coupling might be the cause for the complex ridges and 3D kinematic asymmetries, which does not seem contradictory to our observational analysis shown in this work from the qualitative perspective . 

Compared to \citet{Wang2020c}, we extend the range from [8, 12] kpc to [8, 15]\,kpc with more detailed analysis, and also find two kinds of ridges. Not only we confirm the previous ridges but also we find more than three ridges. Although the coupling mechanism as mentioned above is more possible, different tracers and populations with different methods and modelling might have different observational evidence for some specific physical mechanisms. Here we attempt to use our observational results to discuss the bar contribution.

Motivated by the Hercules origin, \citet{Monari2019a,Monari2019b} proposed there were at least six ridges related to the bar in the local action space using modelling work, which is partly supported by the recent observational work with the same slope of $-$8 km s$^{-1}$ kpc$^{-1}$ \citep{2022ApJ...936L...7C}. Recently, the intriguing break in the $V_{Z}$ and $L_{Z}$ plane was firstly discovered in \citet{Antoja2021} using EDR3 dataset, which is showing a bi-modality for stars with large angular momentum moving vertically upwards and meanwhile for stars with slightly lower angular momentum moving preferentially downwards. These discoveries can be seen as the ridges pattern/projection in the specific phase-space, the break location is at around 2750 kpc km s$^{-1}$ which is corresponding to their new ridge and also this work. With the coming of the Gaia DR3, \citet{2022arXiv220712977D} has revealed a similar break and suggest that it might be originated from the bar's resonance (resonance like feature), but the spiral arms and satellites interaction are not ruled out at all. As shown in this work, the clear ridge features of the north and south phase-space are found, especially for the vertical motions and asymmetries. These might not be explained well by the Milky Way bar dynamics as mentioned in \citep{Wang2020a, Monari2019b,2022MNRAS.516.4988M}, due to the bar should mainly affect the in-plane asymmetries almost equally in the north-south sphere and even for the almost not possible vertical contribution, the bar contribution should also be similar in both sides. So here we suggest that bar should not be enough to explain the ridge properties including the break found in the $V_{Z}$ and $L_{Z}$ plane.

\begin{figure}
  \centering
  \includegraphics[width=0.45\textwidth,height=0.6\textwidth]{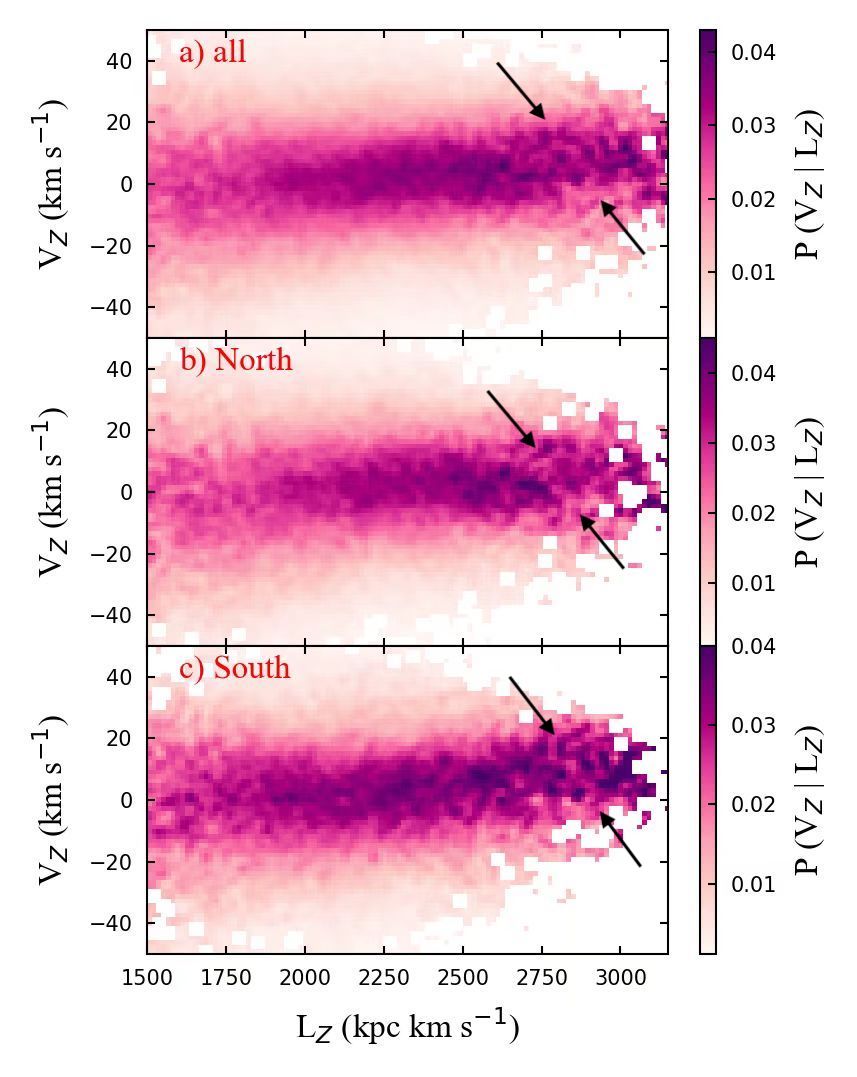}
  \caption{Normalised density in the vertical velocity and angular momentum plane. Top one is the entire sample, middle one is sample distribution in the north side, and bottom one is sample in the south side. Two arrows in the three panels are showing the break discovered in the recent works, especially for the middle one.}
  \label{break}
\end{figure}

\begin{figure}
  \centering
  \includegraphics[width=0.45\textwidth,height=0.6\textwidth]{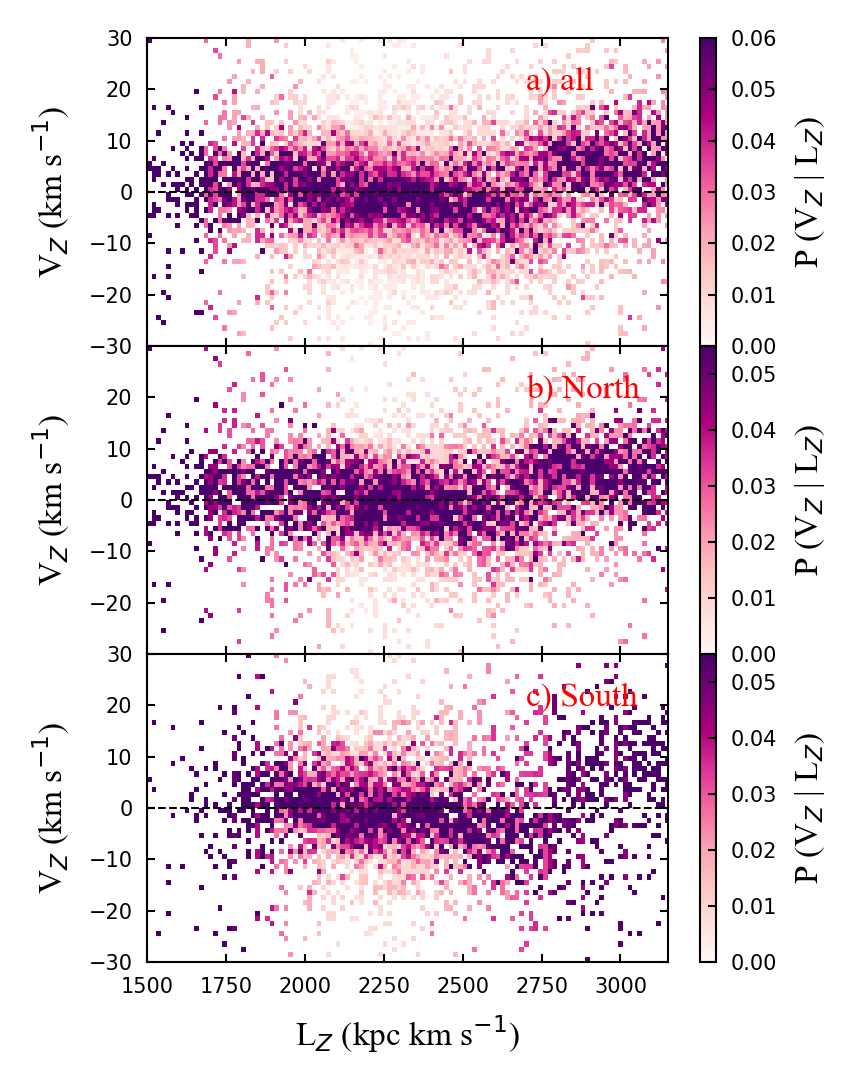}
  \caption{Similar to Fig. \ref{break} but for young OB type stars. Normalised density in the vertical velocity and angular momentum plane. Top one is the entire sample, middle one is sample distribution in the north side, and bottom one is sample in the south side. Three panels are showing the break discovered in the recent works, which is far clearer and stronger than RGB sample.}
  \label{OBbreak}
\end{figure}

Furthermore, we investigate the break in the $V_{Z}$ and $L_{Z}$ plane by using our RGB sample. As shown in Fig. \ref{break}, we can see a break marked with arrows in the top two panels (full sample and north side), especially in the middle one. The angular momentum of the break is about 2900 kpc km s$^{-1}$ (R6), not only is it consistent with the \citet{Antoja2021} (See Sec.3.1), but also is it similar to the recent Gaia DR3 RGB patterns \citep{2022arXiv220712977D}. Note that the RGB pattern for the break is relatively weaker than the younger populations such like Cepheids which were also mentioned in the same recent DR3 work. Here we also use 13534 OB stars from LAMOST \citep{Yu2021} to compare in more details above and below the mid-plane. It is illustrated in Fig. \ref{OBbreak} that the OB break is far clearer than RGB sample, although we note there might be lack of sampling in the south and beyond the break location, but the break is still detected. By focusing on the pattern before the break location, we can also find the south distribution is more inclined than that in the north so that we can say there are differences above and below the plane. In the meantime, that DR3 work also shows that the Cepheid sample has important distinctions with respect to the older RGB and RVS sample, which suggest that the Cepheid gap is distinct from the break in \citet{Antoja2021} and \citet{2022MNRAS.516.4988M}. Then it is pointing that there are some fully open questions for the flipped orientation in action space and young population new gap (2950 kpc km s$^{-1}$) which is very similar to our R6 ridge in consideration of the solar values difference and small relative errors. However, in our work, we surprisingly find that the break is much weaker in the south side for RGB sample (bottom panel of Fig. \ref{break}), that might be caused by the LAMOST sampling and selection effects. Actually the north and south differences are also very clearly discovered in \citet{2022MNRAS.516.4988M}, which might be hard to be explained by the bar for the north-south difference. 

As mentioned above, it is not expected if the break in the ridge patterns is fully contributed by the bar dynamics. In addition, \citet{Antoja2021} also pointed out the vertical height and velocity were not perfect one-to-one correlation (See their Fig. 15), that is also hard  for us to imagine how the bar resonance could explain this phenomenon. Meanwhile, the older population is weaker than the younger population for the break, which might also not be consistent with the influence  caused by the secular evolution of the bar. Moreover, the difference between OB stars and Main-Sequence-Turn-Off stars (MSTO) for the intriguing break around 11$-$12 \,kpc has also been confirmed in our recent work in \citet{lixiang2022}, which is used to discuss warp properties. So it is reasonable to ask whether or not the break is caused by the warp dynamics in consideration of the $V_{Z}$ break and younger populations are stronger than old populations \citep{Chrob2022, Wang2020b}, the origin of the warp could possibly be gas direct infall onto the disk and Sgr interactions (\citet{Lopez2019review} and reference therein). Recently, \citet{2022arXiv221011964S} have proposed that the mutiple-spiral arm can  lead to such a gap by also taking accounts of warp together. Here we don't plan to go more details about the intriguing break for the moment, but we will surely investigate more to deepen the understanding about this feature in the future.

\subsubsection{Ridge related to the rotational velocity?}
\begin{figure}
  \centering
  \includegraphics[width=0.45\textwidth,height=0.6\textwidth]{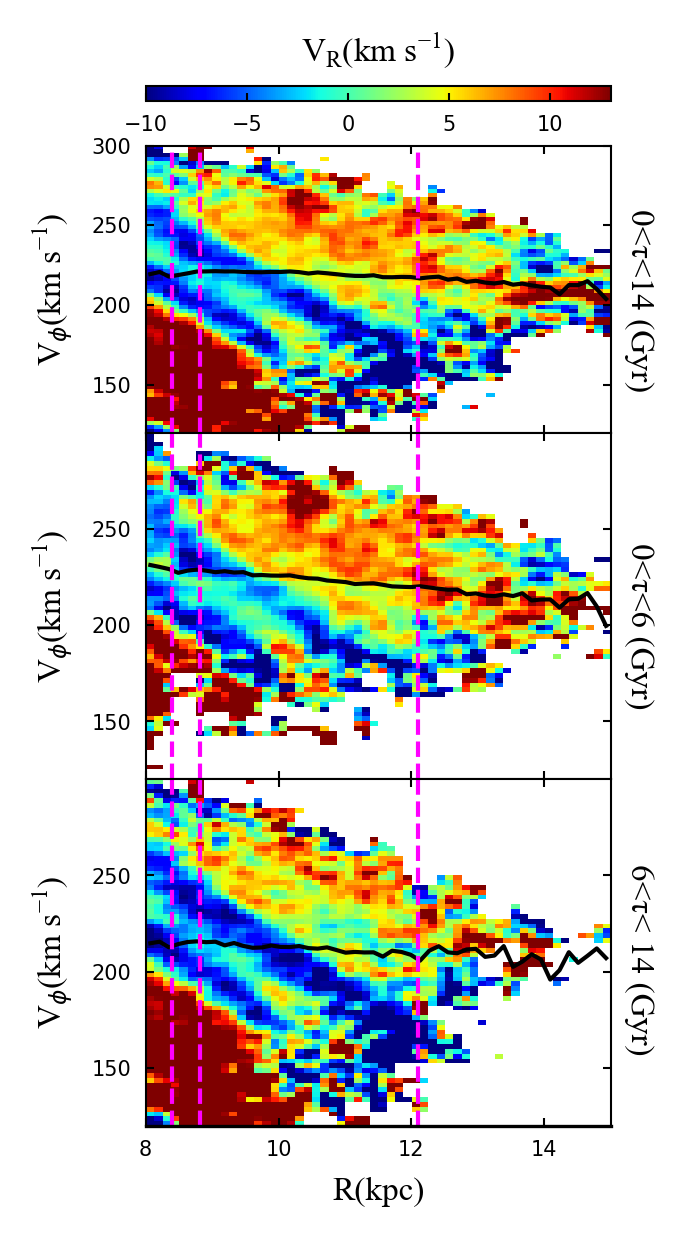}
  \caption{Radial velocity distribution on the $R$ and $V_\phi$ plane and the rotational (azimuthal) velocity is overlapped with black curve. Note that the rotation curve has slight wobble near the ridge region indicated by the vertical dashed lines.}
  \label{rcridge}
\end{figure}

Based on the simulation, \citet{Medina2019,Medina2020} found that the ridge structure has a certain impact on the rotation curve of the Milky Way galaxy such like  bumps and wiggles, which might make the rotation curve swing and fluctuate on the plane near the ridge location. Recently, focusing on the three-dimensional kinematics and age distribution of the open cluster population, \citet{Tarricq2021} found that the ridge structure can also be observed by using this tracer, and the rotation curve unveiled by the open clusters also demonstrates a small swing and rise near the ridge. Similarly, we also present the analysis for the influence of diagonal ridge on the azimuthal/rotational velocity curve. As shown in Fig. \ref{rcridge}, three ridges (R2, R3 and R5) shown as fuchsia dashed lines might have influence on the black rotational curve which has some fluctuations. According to current results, we are consistent with the results of \citet{Medina2019,Medina2020} and more importantly, they also found a total of six ridge structures in the [8,15] \,kpc range with simulations focusing on the imprint of arms and bars on the rotation curve. 

In short, combining the complex and abundant ridge patterns, asymmetrical properties, vertical velocity asymmetries, the break features, and rotation curve signals related to the ridge. We speculate that the bar should not be enough to explain the ridge physics including the break, perhaps Sagittarius dwarf galaxy, spiral arms and warp, etc., are also important contributions for this structure formation. So maybe the coupling mechanisms are more possible, however, it is clear that this is a qualitative discussion, we are still far from the truth of the ridge, more works about observations and simulations are needed.

\section{Conclusions}

In this work, we adopt the RGB sample obtained by cross matching LAMOST and Gaia to investigate the chemical-kinematical-dynamical properties of the well-known ridge structure in the outer disc from 8 to 15\,kpc. We find/confirm there are six ridges in the $R$$-$$V_\phi$ plane and present the time-tagging onto these. Not only in the kinematical space we find the ridge features but also we detect some similar signals in the chemistry space (with gradient) and mass pattern. North-South comparisons are also shown that clear asymmetrical ridge features are almost everywhere such like vertical motions, chemistry and mass. Moreover, the new break feature found in recent Gaia EDR3 work is also confirmed and the possible influence of ridge on the rotation curve (azimuthal velocity) is also unveiled here. 

Furthermore, two kinds of ridges are confirmed again with more physical details. In combination with the ridge kinematic properties, 3D velocity asymmetries, vertical bulk motions, break, rotation curve signals related to ridges and some recent modelling progress. We find that bar resonance/dynamics should not be enough to explain these observational evidence especially for the break, so we speculate that coupling mechanisms with internal and external perturbations are more possible. In the future work, we will make full use of more precise dataset and simulated catalog to push more for this fascinating diagonal ridges topic, in particular the break.

\section*{Acknowledgements}

We would like to thank the anonymous referee for his/her very helpful and insightful comments. We acknowledge the National Key R \& D Program of China (Nos. 2021YFA1600401 and 2021YFA1600400). HFW acknowledges the support from the project “Complexity in self-gravitating systems” of the Enrico Fermi Research Center (Rome, Italy)  and science research grants from the China Manned Space Project with NO. CMS-CSST-2021-B03, CMS-CSST-2021-A08. L.Y.P is supported by the National Natural Science Foundation of China (NSFC) under grant 12173028, the  Chinese Space Station Telescope project: CMS-CSST-2021-A10, the Sichuan Science and Technology Program (Grant No. 2020YFSY0034), the Sichuan Youth Science and Technology Innovation Research Team (Grant No. 21CXTD0038), Major Science and Technology Project of Oinghai Province (Grant No. 2019-ZJ-A10), and the Innovation Team Funds of China West Normal University (Grant No. KCXTD2022-6). TTG acknowledges partial financial support from the Australian Research Council (ARC) through an Australian Laureate Fellowship awarded to Joss Bland-Hawthorn. H.F.W. is enthusiastic  for the plan ``Mapping the Milky Way (Disk) Population Structures and Galactoseismology (MWDPSG) with large sky surveys" in order to establish a theoretical framework in the future to unify/partly unify the global picture of the disk structures and origins with a possible comprehensive distribution function. The Guo Shou Jing Telescope (the Large Sky Area Multi-Object Firber Spectroscopic Telescope, LAMOST) is a National Major Scientific Project built by the Chinese Academy of Sciences. Funding for the project has been provided by the National Development and Reform Commission. LAMOST is operated and managed by National Astronomical Observatories, Chinese Academy of Sciences. This work has also made use of data from the European Space Agency (ESA) mission {\it Gaia} (\url{https://www.cosmos.esa.int/gaia}), processed by the {\it Gaia} Data Processing and Analysis Consortium (DPAC, \url{https://www.cosmos.esa.int/web/gaia/dpac/consortium}). Funding for the DPAC has been provided by national institutions, in particular the institutions participating in the {\it Gaia} Multilateral Agreement.


\end{document}